\documentclass[conference]{IEEEtran}

\usepackage{color}
\usepackage{amssymb}
\usepackage{rotating}
\usepackage[T1]{fontenc}
\usepackage{lscape}
\usepackage{dblfloatfix}
\usepackage{xcolor}
\usepackage{subcaption}

\ifCLASSINFOpdf
\else
 
\fi

\begin{document}
%
\title{Efficient multi-descriptor fusion for non-intrusive appliance recognition}


\author{\IEEEauthorblockN{Yassine Himeur, Abdullah Alsalemi, Faycal Bensaali}
\IEEEauthorblockA{\textit{Department of Electrical Engineering,} \\
\textit{Qatar University, Doha, Qatar}\\
Email: yassine.himeur@qu.edu.qa, a.alsalemi@qu.edu.qa, \\
 f.bensaali@qu.edu.qa}

\and
\IEEEauthorblockN{Abbes Amira}
\IEEEauthorblockA{\textit{Institute of Artificial Intelligence,} \\
\textit{De Montfort University, Leicester, United Kingdom}\\
Email: abbes.amira@dmu.ac.uk}
}


\maketitle

\begin{abstract}
Consciousness about power consumption at the appliance level can assist user in promoting energy efficiency in households. In this paper, a superior non-intrusive appliance recognition method that can provide particular consumption footprints of each appliance is proposed. Electrical devices are well recognized by the combination of different descriptors via the following steps: (a) investigating the applicability along with performance comparability of several time-domain (TD) feature extraction schemes; (b) exploring their complementary features; and (c) making use of a new design of the ensemble bagging tree (EBT) classifier. Consequently, a powerful feature extraction technique based on the fusion of TD features is proposed, namely fTDF, aimed at improving the feature discrimination ability and optimizing the recognition task. An extensive experimental performance assessment is performed on two different datasets called the GREEND and WITHED, where power consumption signatures were gathered at 1 Hz and 44000 Hz sampling frequencies, respectively. The obtained results revealed prime efficiency of the proposed fTDF based EBT system in comparison with other TD descriptors and machine learning classifiers.

\end{abstract}

\IEEEpeerreviewmaketitle

\section{Introduction}

Energy saving and the reduction of carbon emission are progressively considered as priorities, not only for individual users, but for decision makers as well \cite{He2019SG,Alsalemi2018IEESyst,Devlin2019}. To that end, research on non-intrusive load monitoring (NILM) and appliance recognition \cite{Welikala2019SG,Andrean2018,Liu2019CE} have been investigated to collect specific appliance consumption statistics at low cost without a request for implementing individual sub-meters. In that respect, capturing individual device energy usage statistics can help detect energy-hungry appliances, biggest energy wasters and  anomalous consumption \cite{Li2019SG,Gaur2018}. Thereby, we can enhance energy demand management and reduce wasted electricity via generating appropriate recommendations for individual users \cite{ALSALEMI2019,Alsalemi2019SBC,Sardinos2019}.

Several characteristic extraction descriptors have been investigated in the state-of-the-art literature. These comprise harmonics and transient features (HTF) \cite{Hart1992}, spectral decomposition \cite{WELIKALA2019}, correlation-based power analytics \cite{ROY2016132}, feature based events detection \cite{Wu2019,Wang2013}, power peaks detection \cite{YAN2019}, wavelet analysis \cite{Gillis2017,Chang2014} and voltage-current trajectory (V-I)\cite{Liu2019SG}. In addition, a comparison of 36 feature descriptors that were explored for appliance recognition is illustrated in \cite{Kahl2017}. In spite of these efforts, developments of robust and efficient non-intrusive appliance recognition systems that can be easily used for in-home energy monitoring, are perceptibly confined. Further, research in that respect, still exposes a gap between academics and industry requirements. Many factors are responsible for this difference including poor recognition accuracy, variability of data from a region to another and the lack of robustness against interferences generated through the measurement campaign \cite{RAFSANJANI2020,CHEN2019252}. 

In this paper, we propose a simple yet effective multi-descriptor fusion that not only provide high performances in device recognition but also support a real-time implementation as it is based on the joint use of various time-domain (TD) feature extraction schemes, which requires significantly lower computational complexity than time-frequency analysis. Accordingly, the proposed fusion of TD features denoted as fTDF (fusion of TD features) can not only derive characteristic profiles from the current segmentation windows, but also combines those with features collected from previous windows to form more efficient appliance recognition. In this respect, the regional variation of the power consumption properties is correlated to the semi-global variation as drawn off using the TD analysis of the past windows. Moreover, in view of achieving a high accuracy, an ensemble bagging tree (EBT) classifier is designed and its performance is then compared in terms of the accuracy and F-score with various other classification models.

\section{Proposed system}
\begin{figure}[t!]
\begin{center}
\includegraphics[width=9cm, height=14cm]{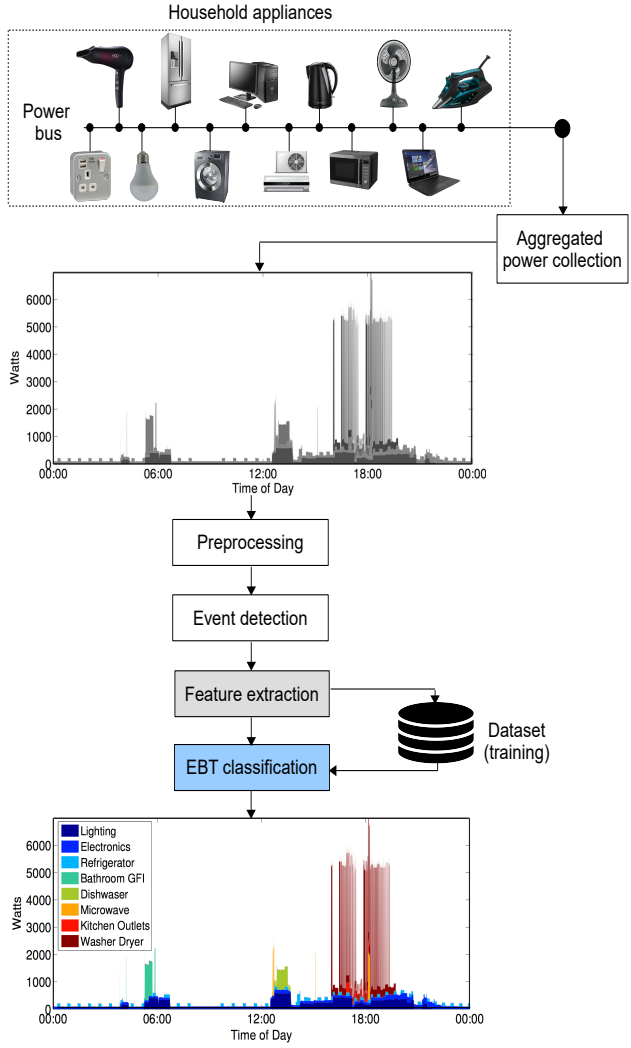}
\end{center}
\caption{Block diagram of the proposed appliance recognition system.}
\label{AppRec}
\end{figure}

A typical block diagram of a non-intrusive appliance recognition system is illustrated in Fig. \ref{AppRec}. 


\subsection{Feature extraction} \label{TDdescriptors}
With a view to describe the TD feature descriptors used under our framework, let us consider a sampled version of the energy consumption signal denominated as: $s[i]$, with $i = 1,2, \cdots, M$, of length $M$ and collected at the sampling frequency $f_{s}$, in order to extract the TD features, a windowing process is applied on $s$ where a window length $N$ is utilized and the TD property $S(k)$ of each window $k$ $(k=1,2, \cdots, K$ and $K$ is the number of extracted windows) is then collected as follows:   

\begin{itemize}
\item Root mean square feature (RMSF)
\begin{equation}
S_{RMS}(k)=\textstyle\sum\limits_{i=1}^{N}\sqrt{\frac{1}{N}(s_{i}^{2})}
\end{equation}

\item Mean absolute deviation feature (MADF)
\begin{equation}
S_{MAD}(k)=\textstyle\sum\limits_{i=1}^{N}\frac{1}{N}\left\vert s_{i}-\mu \right\vert 
\end{equation}
where $\mu$ represent the central tendency,

\item Integrated absolute magnitude feature (IAMF)
\begin{equation}
S_{IAMF}(k)=\frac{1}{N}\textstyle\sum\limits_{i=1}^{N}\frac{s_{i}^{2}}{2}\mathrm{sgn}%
(s_{i})+\mu 
\end{equation}

\item Zero crossing feature (ZCF)
\begin{equation}
S_{ZC}(k)=\textstyle\sum\limits_{i=2}^{N}\left\vert \textnormal{sgn}(s_{i})-\textnormal{sgn}(s_{i-1})\right\vert 
\end{equation}

\item Waveform length feature (WLF)
\begin{equation}
S_{WL}(k)=\log \left( \textstyle\sum\limits_{i=1}^{N-1}\left\vert
s_{i+1}-s_{i}\right\vert \right) =\log \left(
\textstyle\sum\limits_{i=1}^{N-1}\left\vert \Delta s_{i}\right\vert \right) 
\end{equation}

\item Slope sign change feature (SSCF)
\begin{equation}
S_{SSC}(k)=\textstyle\sum\limits_{i=2}^{N-1}f[(s_{i}-s_{i-1})\times (s_{i}-s_{i+1})]
\end{equation}
where 
\begin{equation}
f(s)=\left\{ 
\begin{array}{cc}
1 & \textnormal{if }s\geq \textnormal{threshold} \\ 
0 & \textnormal{otherwise \ \ \ \ \ \ \ \ }%
\end{array}%
\right. 
\end{equation}

\item Auto-regressive feature (ARF)
\begin{equation}
S_{ARF}(k)=\textstyle\sum\limits_{i=1}^{N}\left(
\sum\limits_{p=1}^{P}a_{p}s_{i-p}+w_{i}\right) 
\end{equation}
where $s_{i-p}$ are the previous samples of the power consumption signal, $w_{i}$ is a white noise and $P$ represents the auto-regressive (AR) model order. Under this framework, $P=15$ is considered.
\end{itemize}

After extracting the features $S(k), k=1,2, \cdots, K$ from the different windows using an overlapping process, these data are concatenated to form the whole feature vector  $S_{F}=[S(1) S(2), \cdots, S(K)]$ of each TD descriptor.

\subsection{Fusion of TD features (fTDF)}
\begin{figure}[t!]
\begin{center}
\includegraphics[width=9.2cm, height=7cm]{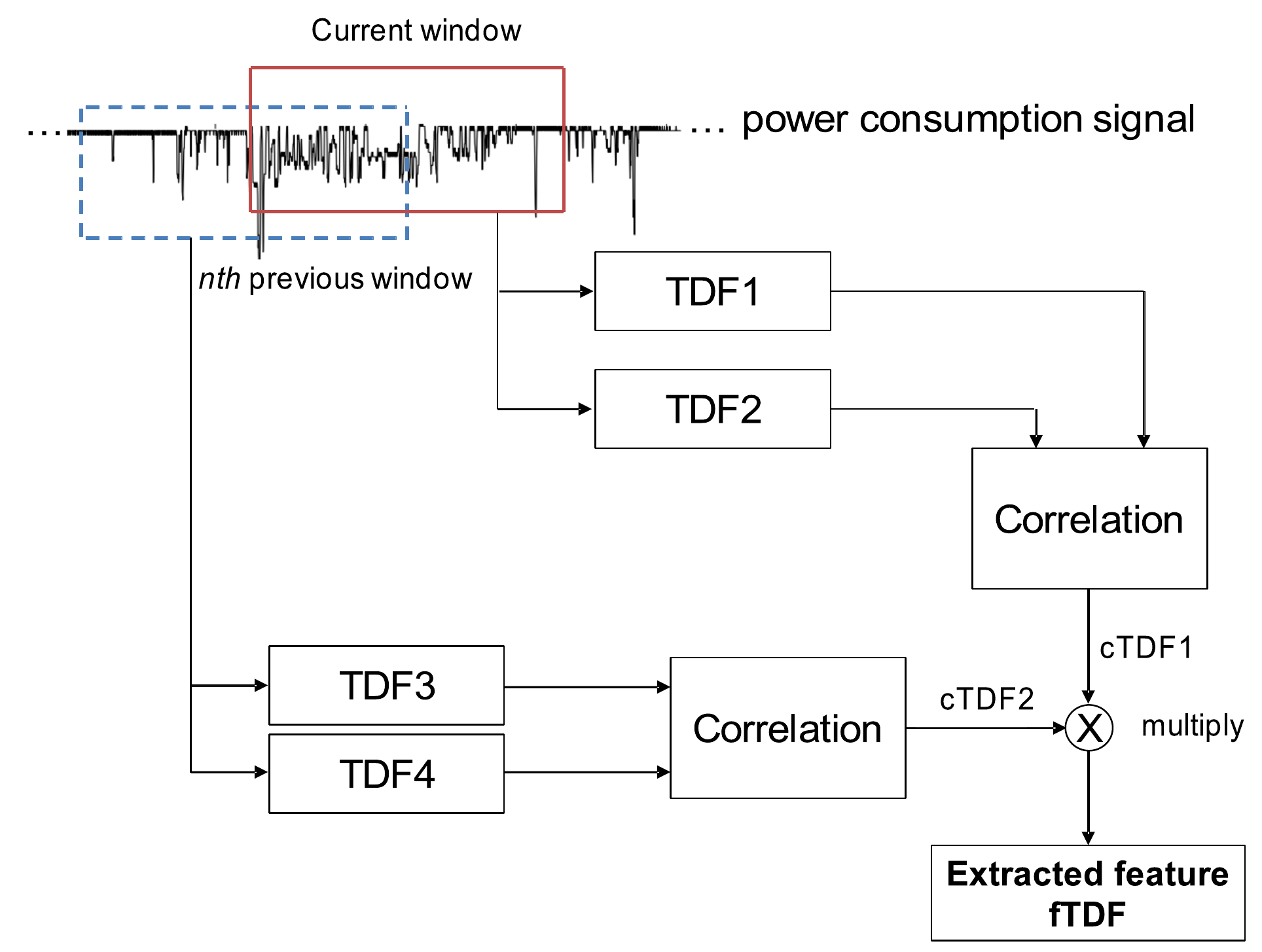}
\end{center}
\caption{Block diagram of the proposed feature extraction based data fusion architecture.}
\label{FusionData}
\end{figure}

As a second contribution,  a data fusion strategy named fTDF is developed, in which the characteristic vector chosen from the current window is fused with the characteristic vectors derived from the $nth$ previous window using four different TD descriptors. In this respect, if the previous characteristic window is from the same appliance category, then the correlation rate will increase. On the flipside, for dissimilar appliance groups, the correlation rate will diminish.  In addition, following the same process, we can correlate with the $2rd$, $3rd$, and other window vectors as well. This experiment demonstrated that not only the first previous windows can be used for correlation, but further apart windows can also be considered ( e.g., the $3th$ previous window vector can be a good alternative since there is no performance improvement after that point). 
The flowchart of the proposed fTDF approach is depicted in Fig. \ref{FusionData},  where the variation of the energy consumption data at each window is correlated to the semi-global variation of the past windows using two different TD descriptors TDF1 and TDF2. The same process is repeated using two other TD descriptors TDF3 and TDF4. Then, the correlation obtained from both scenarios is fused using a multiplication process to form the novel feature vector. 

\subsection{Ensemble bagging tree (EBT) classifier}
The EBT model is a powerful classifier that did not receive its merit in
practice. The importance of EBT comes from the fact that it can achieve a high
classification performance by using a fusion of various weak classifiers. If
we consider a prediction function $g(y,K)$ for predicting a category label $%
c\in \{1,2,3...C$, where $C$ represents the whole number of categories).

\begin{equation}
q(c\mid y)=P(g(y,K)=c)
\end{equation}
and the whole probability of correct classification can be given as

\begin{equation}
w_{c}=[\sum\limits_{c}q(c\mid y)P(c\mid y)]P_{y}d_{y}    \label{eq0}
\end{equation}
where $q(c\mid y)$ represents the conditional probability for the forecasted
group, hence, the total accuracy  can be reached using the formula of Eq. \ref{eq0}, $Py(y)$ outlines the probability distribution of y. 
In addition, if $O$ is the original set of all inputs $y$ where $\varphi $ is order-correct, the correct classification rate can be estimated as

\begin{equation}
\begin{array}{c}
w_{A}=\int_{y\in O}\max P(c\mid y)Py(y) \\ 
+\int_{y\in O^{\prime }}\left[ \sum_{c}I(\varphi _{A}(y)=c)P(c\mid y)Py(y)%
\right] 
\end{array}
\label{eq1}
\end{equation}
where $I(\cdot )$ represents the indicator function. From Eq. \ref{eq1}, it can be drawn that for an order-correct $\varphi $ for the inputs $y$, the accurate classification ratio is a long way of being optimal, while $\varphi _{A}$ is optimal. Moreover, we can also deduce that  even if a classifier is good (i.e., order-correct for nearly all the inputs $y$), it can be considered nearly optimal when it is constituted of multiple weak classifiers.  

\section{Experimental results}
\subsection{Datasets description}
Two datasets are used in this framework to validate the proposed appliance recognition system. They are the GREEND \cite{GREEND2014} and WHITED \cite{WHITED2016} collected at low and high frequency sampling rates, respectively. The WHITED includes the power consumption signatures of the device start-ups collected from up 110 electrical devices, categorizing up to 47 appliance groups. For each appliance category, a set of power consumption fingerprints is collected from different appliance manufacturers and all of them are gathered at a sampling rate of 44000 Hz. In this framework, we use 11 appliance categories to validate the proposed system. The GREEND contains the electricity consumption signatures of several domestic appliances collected at a sampling frequency of 1 Hz from 8 households in Italy and Austria. To validate the proposed system, we use energy usage footprints gathered from typical house. Table \ref{WHITED} depicts the electrical devices used to collect load consumption signatures and the monitoring period for both the WHITED and GREEND.

\begin{table}[t!]
\caption{Description of monitored appliances on both the WHITED and GREEND datasets}
\label{WHITED}
\begin{center}

\begin{tabular}{lll|lll}
\hline
\multicolumn{3}{c|}{WHITED} & \multicolumn{3}{|c}{GREEND} \\ \hline
Tag & Apliance  & \# mesured  & Tag & Appliance & \# mesured \\ 
& category & apliances &  & Category & \multicolumn{1}{r}{days} \\ \hline
1 & Modem/receiver & \multicolumn{1}{r|}{20} & 1 & Coffee machine & 
\multicolumn{1}{r}{242} \\ 
2 & CFL & \multicolumn{1}{r|}{20} & 2 & Radio & \multicolumn{1}{r}{242} \\ 
3 & Charger & \multicolumn{1}{r|}{30} & 3 & Fridge / freezer & 
\multicolumn{1}{r}{240} \\ 
4 & Coffee machine & \multicolumn{1}{r|}{20} & 4 & Dishwasher & 
\multicolumn{1}{r}{242} \\ 
5 & Drilling machine & \multicolumn{1}{r|}{20} & 5 & kitchen lamp & 
\multicolumn{1}{r}{242} \\ 
6 & Fan & \multicolumn{1}{r|}{30} & 6 & TV & \multicolumn{1}{r}{242} \\ 
7 & Flatron & \multicolumn{1}{r|}{20} &  &  &  \\ 
8 & LED ight & \multicolumn{1}{r|}{20} &  &  &  \\ 
9 & Kettle & \multicolumn{1}{r|}{20} &  &  &  \\ 
10 & Microwave & \multicolumn{1}{r|}{20} &  &  &  \\ 
11 & Iron & \multicolumn{1}{r|}{20} &  &  &  \\ \hline
\end{tabular}

\end{center}
\end{table}

MATLAB 2018a is employed to carry out the analysis. A segmentation based on an overlapping rate of 1/4 the window size is utilized for all the descriptors during this study. The  proposed fTDF approach is designed using the fusion of four TD decriptors, including MADV, IAVF, RMSF and WLF.  With a view of evaluating the performance of the proposed fTDF, the obtained results' accuracy (acc) and F-score (F-scr) are compared to those extracted from the various TD descriptors described in section \ref{TDdescriptors}.

\begin{figure}[t!]
\begin{center}
\includegraphics[width=7.6cm, height=5.1cm]{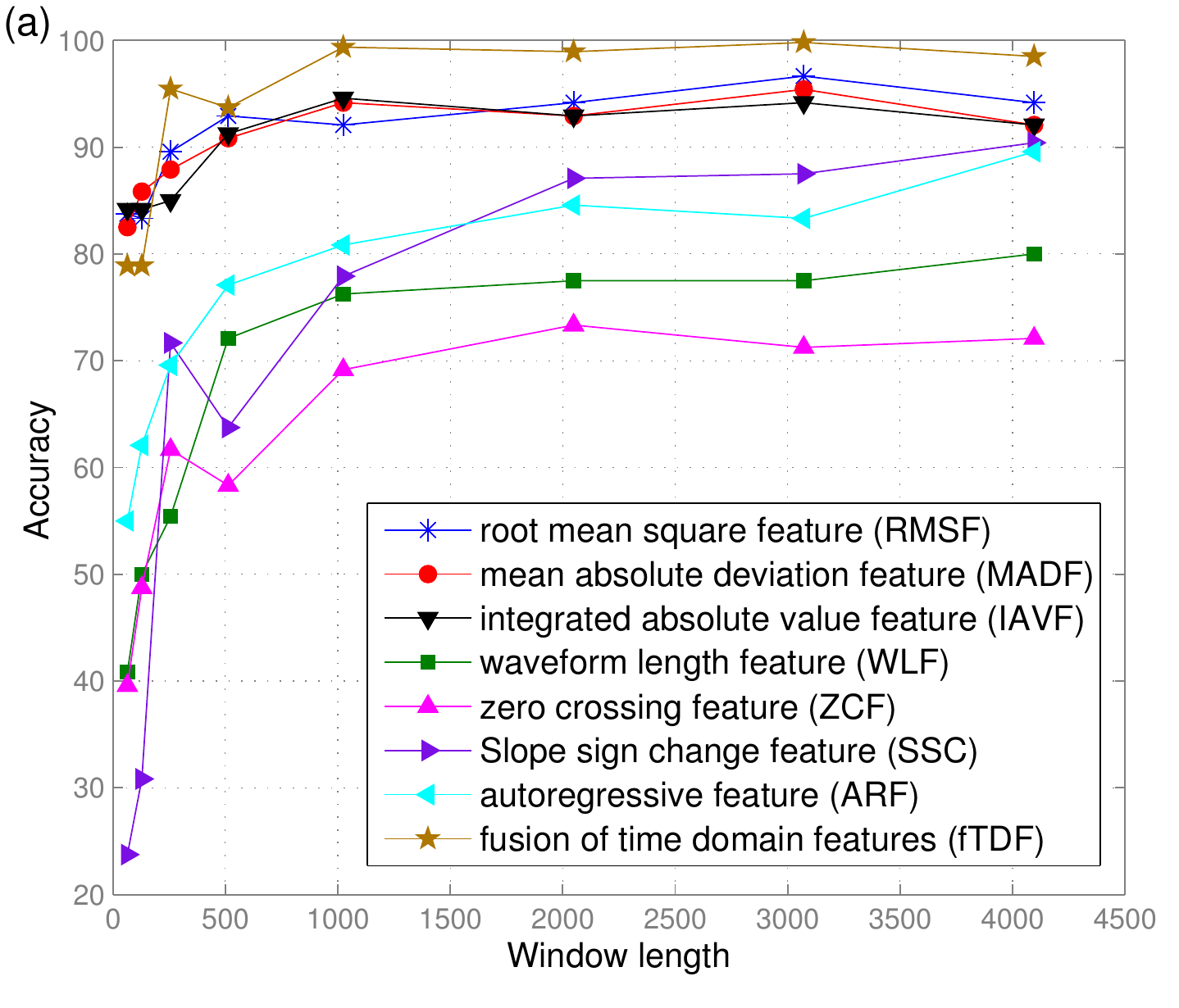}
\includegraphics[width=7.6cm, height=5.1cm]{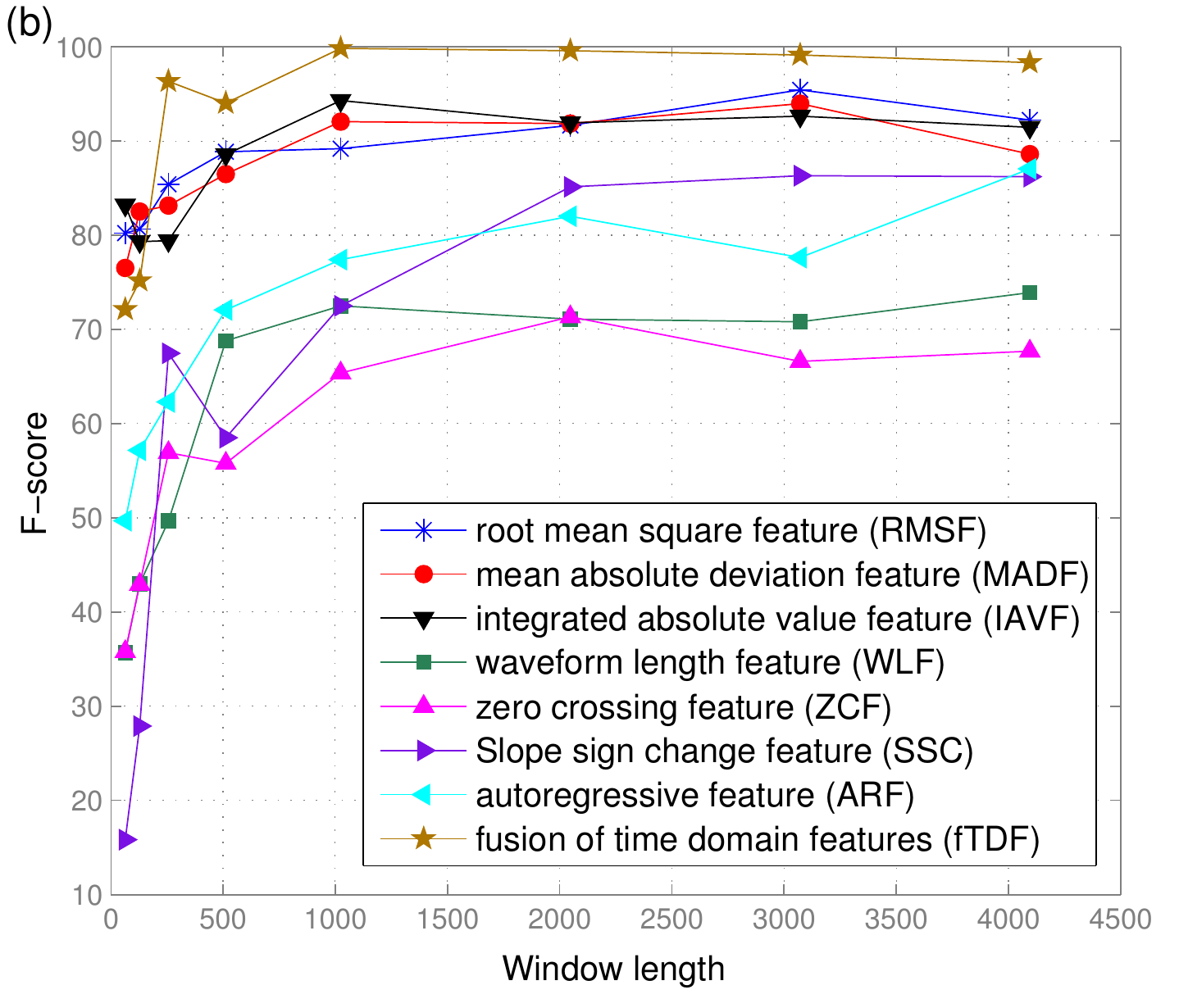}
\end{center}
\caption{Performances variation according to the window length for the WHITED: a) accuracy and b) F-score.}
\label{WHITEDper}
\end{figure}

\begin{figure}[t!]
\begin{center}
\includegraphics[width=7.6cm, height=5.1cm]{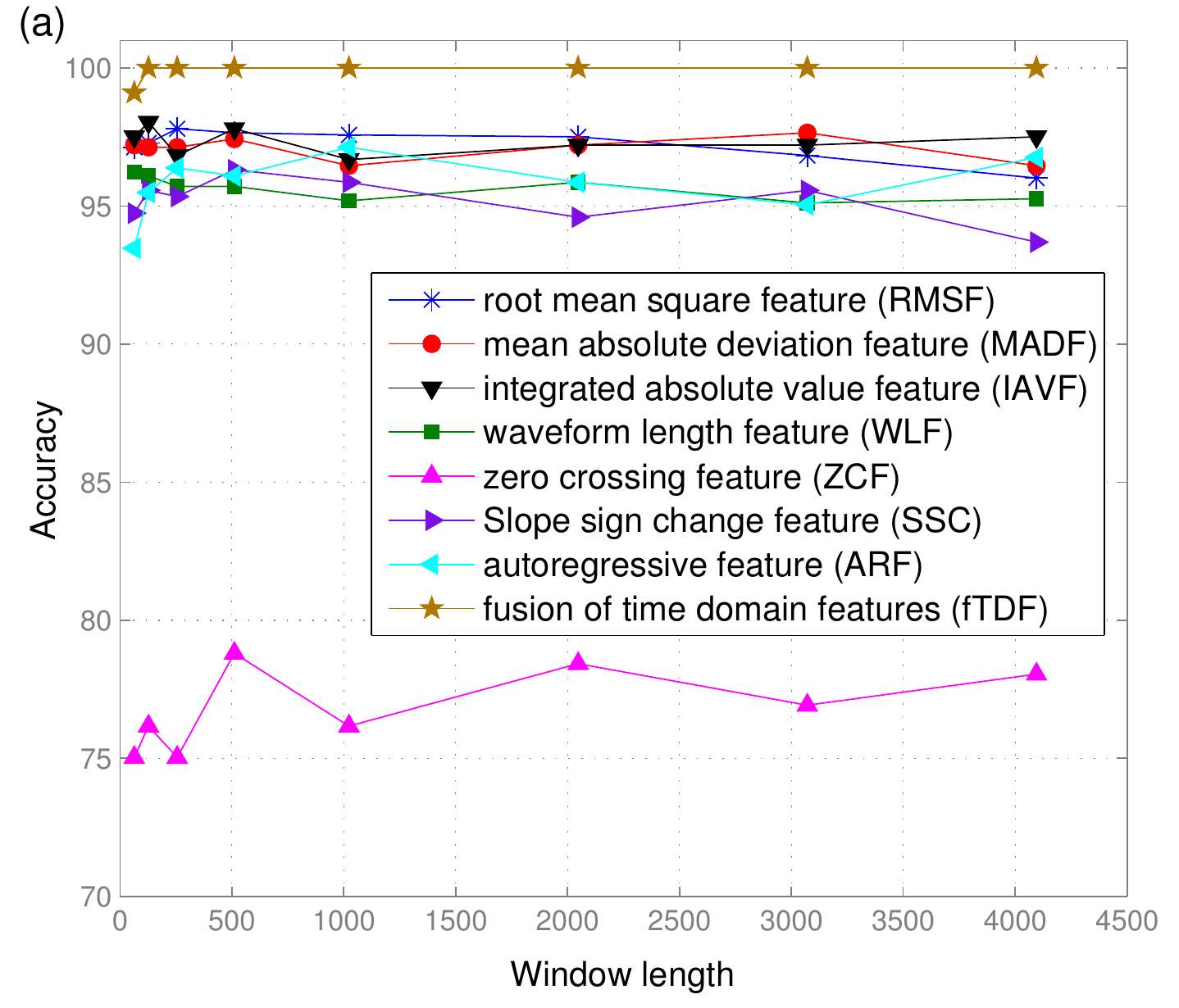}
\includegraphics[width=7.6cm, height=5.1cm]{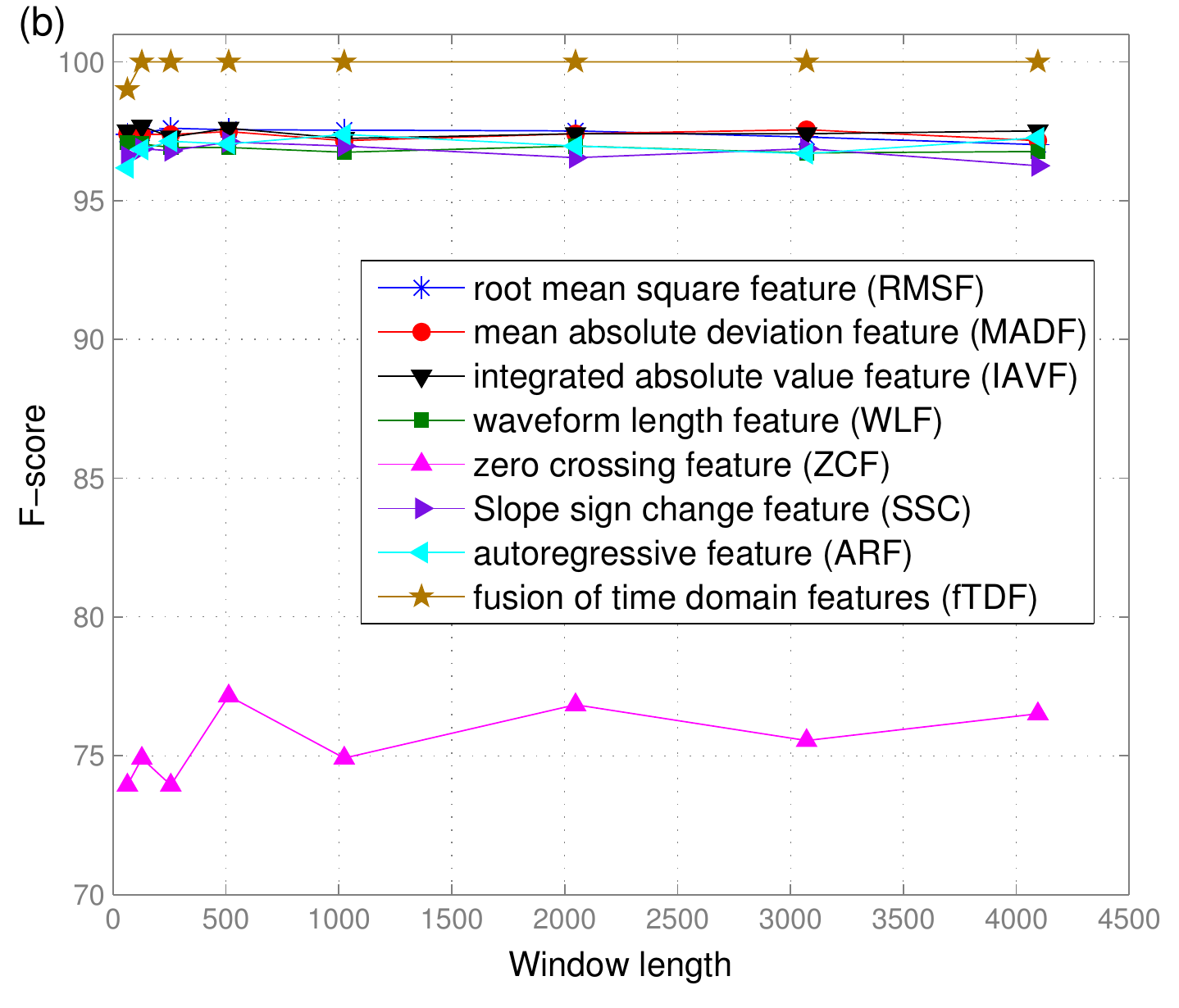}
\end{center}
\caption{Performances variation according to the window length for the GREEND: a) accuracy and b) F-score.}
\label{GREENDper}
\end{figure}

Fig. \ref{WHITEDper} and Fig. \ref{GREENDper} portray the comparison of the accuracy and F-score performances of fTDF against the other TD feature extraction schemes, according to a window size ranging from 64-4096 for both the GREEND and WHITED datasets. The obtained results show the superiority of the proposed fTDF in both cases. The best performances can be obtained for a window length of 3072. This can be justified by the fact that through aggregating several weak descriptors using the proposed fusion architecture, a powerful descriptor is designed that can improve feature discrimination ability and thus optimize the classification process. Therefore, this is the main advantage of the this fusion process. Furthermore, the slight difference between the results collected from the WHITED and GREEND is due to the fact that these datasets are quite different. The first collects the power consumption signatures of each device for a very short duration (5 sec) while the second one gathers energy usage footprints of each appliance for the whole day.

\subsection{Comparison with other classifiers}
We perform a performance comparison of the EBT classifier versus other machine learning models including support vector machine (SVM), deep neural networks (DNN), K-nearest neighbors (KNN), decision tree (DT) operating by reference to different classification parameters and when the fTDF is considered as well. Table \ref{classComp} illustrates the accuracy and F-score outputs obtained for both the GREEND and WHITED datasets where a window length of 3072 is considered. It is clearly shown that the EBT outperformed the other classifiers with respect to the accuracy and F-score.

\begin{table}[t!]
\caption{Accuracy and F-score of the EBT model compared with other classifiers using fTDF}
\label{classComp}
\begin{center}

\begin{tabular}{l|c|c|c|c|c}
\hline
\textbf{ML } & \textbf{Classifier} & \multicolumn{2}{|c}{\textbf{GREEND}} & 
\multicolumn{2}{|c}{\textbf{WHITED}} \\ \cline{3-6}\cline{3-5}
\textbf{classifier} & \textbf{\ parameters} & \textbf{acc} & \textbf{F-scr}
& \textbf{acc} & \textbf{F-scr} \\ \hline
{\small SVM} & {\small Linear Kernel} & 83.41 & 82.74 & 96.5 & 96.14 \\ 
\hline
{\small SVM} & {\small Quadratic kernel} & 77.45 & 72.67 & 97.16 & 97.3 \\ 
\hline
{\small SVM} & {\small Gaussian kernel} & 83.86 & 82.99 & 88.75 & 85.75 \\ 
\hline
{\small KNN} & {\small K=1, Euclidean dist} & 96.53 & 96.54 & 83.75 & 78.77
\\ \hline
{\small KNN} & {\small K=10, Weighted} & 95.78 & 95.77 & 82.91 & 76.8 \\ 
& {\small Euclidean dist} &  &  &  &  \\ \hline
{\small KNN} & {\small K=10, Cosine dist} & 93.13 & 93.09 & 81.25 & 74.13 \\ 
\hline
{\small DT} & {\small Fine, 100 splits} & 93.96 & 93.95 & 81.25 & 77.93 \\ 
\hline
{\small DT} & {\small Medium, 20 splits} & 94.64 & 94.66 & 82.5 & 79.7 \\ 
\hline
{\small DT} & {\small Coarse, 4 splits} & 62.89 & 53.81 & 39.58 & 28.94 \\ 
\hline
{\small DNN} & {\small 50 hidden layers} & 97.11 & 96.93 & 96.87 & 96.61 \\ 
\hline
\textbf{EBT} & \textbf{30 learners, 42 k} & \textbf{100} & \textbf{100} & 
\textbf{99.69} & \textbf{99.47} \\ 
& \textbf{\ splits} &  &  &  &  \\ \hline
\end{tabular}

\end{center}
\end{table}

\section{Conclusion}

In this framework, a robust non-intrusive appliance system based on a multi-descriptor fusion was proposed. First, various TD descriptors were explored individually to verify their corresponding performances. Second, based on the obtained results, a multiple TD feature fusion technique denoted as fTDF was introduced. In this proposed technique, the correlation vectors extracted from four TD descriptors using an improved windowing process were then combined. This helped mitigating the limitation of each one and also improving overall feature discrimination ability. Third, the new EBT design was introduced to classify the fTDF patterns extracted from two realistic datasets. The attained results showed improved results compared to various TD descriptors and other classifiers. Future works will aim to analyze appliance specific data collected from the fTDF based EBT system, detect abnormal usage and then generate appropriate recommendations to enhance energy efficiency.

\section*{Acknowledgements}\label{acknowledgements}
This paper was made possible by National Priorities Research Program (NPRP) grant No. 10-0130-170288 from the Qatar National Research Fund (a member of Qatar Foundation). The statements made herein are solely the responsibility of the authors.

\end{document}